\documentclass[aps,prd]{revtex4-1}
\usepackage{graphicx}
\usepackage{amsmath}
\usepackage{array}
\usepackage{bm}
\usepackage{amssymb}
\usepackage{amsfonts}

\begin{document}

\title{A large-$N_c$ PNJL model with explicit Z$_{N_c}$ symmetry}

\author{Fabien \surname{Buisseret}}
\email[E-mail: ]{fabien.buisseret@umons.ac.be}
\affiliation{Service de Physique Nucl\'{e}aire et Subnucl\'{e}aire,
Universit\'{e} de Mons--UMONS,
Acad\'{e}mie universitaire Wallonie-Bruxelles,
Place du Parc 20, B-7000 Mons, Belgium;\\ \\
Haute Ecole Louvain en Hainaut (HELHa), Chauss\'ee de Binche 159, B-7000 Mons, Belgium}

\author{Gwendolyn \surname{Lacroix}}
\email[E-mail: ]{gwendolyn.lacroix@umons.ac.be}
\thanks{F.R.S.-FNRS Research Fellow}
\affiliation{Service de Physique Nucl\'{e}aire et Subnucl\'{e}aire,
Universit\'{e} de Mons--UMONS,
Acad\'{e}mie universitaire Wallonie-Bruxelles,
Place du Parc 20, B-7000 Mons, Belgium}

\date{\today}

\begin{abstract}
A PNJL model is built, in which the Polyakov-loop potential is explicitly Z$_{N_c}$-symmetric in order to mimic a Yang-Mills theory with gauge group SU($N_c$). The physically expected large-$N_c$ and large-$T$ behaviours of the thermodynamic observables computed from the Polyakov-loop potential are used to constrain its free parameters. The effective potential is eventually U(1)-symmetric when $N_c$ is infinite. Light quark flavours are added by using a Nambu-Jona-Lasinio (NJL) model coupled to the Polyakov loop (the PNJL model), and the different phases of the resulting PNJL model are discussed in 't Hooft's large-$N_c$ limit. Three phases are found, in agreement with previous studies resorting to effective approaches of QCD. When the temperature $T$ is larger than some deconfinement temperature $T_d$, the system is in a deconfined, chirally symmetric, phase for any quark chemical potential $\mu$. When $T<T_d$ however, the system is in a confined phase in which chiral symmetry is either broken or not. The critical line $T_\chi(\mu)$, signalling the restoration of chiral symmetry, has the same qualitative features than what can be obtained within a standard $N_c=3$ PNJL model.  
\end{abstract}

\maketitle 

\section{Introduction}

The structure of the QCD phase diagram is intimately related to our understanding of fundamental features of QCD, like for example confinement dynamics and chiral symmetry breaking, and to their interplay with in-medium effects like a nonzero temperature or quark density. This is the reason why a lot of effort is devoted to study this field, either on the theoretical side, to which the present work belongs, or on the experimental side through heavy-ion-collision experiments. Among the various effective frameworks used to study the QCD phase diagram (see \textit{e.g.} the review~\cite{review}), we will mostly focus on two of them: Polyakov-loop effective models for the pure gauge part of QCD, and the Nambu-Jona-Lasinio (NJL) model for the quark part. 

The Polyakov loop is defined as
\begin{equation}\label{L}
L(T, \vec x)= P\, {\rm e}^{i\, g\int^{1/T}_0d\tau A_0(\tau, \vec x)},
\end{equation}
in which $P$ is the path-ordering, $g$ the strong coupling constant, $A_0=A_0^a\, T_a$ the temporal component of the Yang-Mills field, $T_a$ the generators of the gauge algebra, and $T$ the temperature. The integral runs on the compactified timelike dimension. The Polyakov loop is a precious tool to study the phase structure of a given Yang-Mills theory since $\left\langle L(T,\vec x) \right\rangle=0 $ $(\neq 0)$ when the theory is in a (de)confined phase~\cite{Polya}. Moreover, gauge transformations belonging to the center of the gauge algebra only cause $L(T,\vec x)$ to be multiplied by an overall factor. That is why it has been conjectured that the confinement/deconfinement phase transition in a Yang-Mills theory with gauge algebra $\mathfrak{g}$ might be related to the spontaneous breaking of a global symmetry related to the center of $\mathfrak{g}$~\cite{sve82}. In the particular case of SU($N_c$), deconfinement might thus be driven by the breaking of a global Z$_{N_c}$ symmetry. The order parameter of the deconfinement phase transition should then be the traced Polyakov loop 
\begin{equation}\label{phi}
\phi=\frac{1}{N_c}{\rm Tr}_c L,
\end{equation}
where the trace ${\rm Tr}_c$ is taken over the colour indices. The thermodynamic properties of pure gauge SU(3) QCD can then be studied by resorting to an effective scalar field theory where the potential energy density is Z$_3$-symmetric, with \textit{e.g.} the form~\cite{Pisa}
\begin{equation}\label{Ud}
U= T^4\, \lambda  \left[ -\frac{b_2(T)}{2}|\phi|^2+ \frac{b_4}{4}\, |\phi|^4+\frac{b_6}{6}(\phi^3+\phi^{*3})\right].
\end{equation}
The real coefficients $b_i$ can be fitted on lattice data. Various applications of this formalism can be found for example in~\cite{Pisa_app}. Note that, in the following, $\phi$ and $L$ will generally be indifferently called Polyakov loop.

The NJL model is based on the Lagrangian~\cite{NJL0}
\begin{equation}\label{njlL}
{\cal L}_{NJL}=\bar q(i\gamma^\mu\partial_\mu-m_q)q+\frac{G}{2}\left[(\bar q q)^2+(\bar q i\gamma_5 \vec \tau q)^2 \right],
\end{equation}
where $q$ is the quark field, $m_q$ the mass matrix, and $\vec \tau$ the Pauli matrices when an SU(2) flavour symmetry is considered. The interaction terms are such that the Lagrangian is chirally symmetric. The NJL model is designed to model chiral symmetry breaking and study many related phenomenological problems; the interested reader may consult the review~\cite{NJL} for more information. In the original NJL model, fermions are not coupled to the gauge field: As shown in \cite{fuku03}, the coupling of this model to the Polyakov loop can be achieved by minimally coupling the quark field to a gauge field of the form  $A_\mu=A_0\, \delta_{\mu0}$, that formally appears as an imaginary quark chemical potential. The so-called PNJL model resulting in this coupling has motivated a lot of studies devoted to the QCD phase diagram~\cite{ratti06,general}, including cases with a nonzero magnetic field~\cite{magnetic} or nonlocal extensions \cite{nonlo,nonlo2}.

\begin{figure}[ht]
\includegraphics[width=8.5cm]{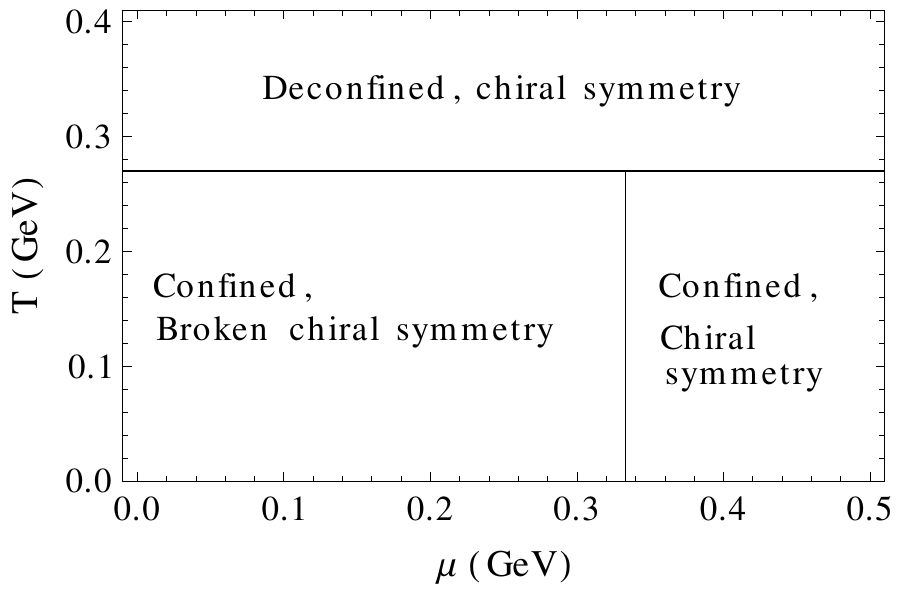}
\caption{Phase diagram obtained by taking the large-$N_c$ limit ($N_c\rightarrow\infty$) of the PNJL model used in~\cite{mcle09}. The solid lines signal first-order phase transitions.}
\label{fig1}
\end{figure}

The phase structure of the PNJL model at arbitrary $N_c$ has been discussed in~\cite{mcle09}, as well as its large-$N_c$ limit. One of the ingredients of this last work is to set $\lambda\propto (N^2_c-1)$ in (\ref{Ud}) so that the gluon potential has the correct scaling in $N_c$. The phase diagram that has been found at large-$N_c$ is given in Fig.~\ref{fig1}. The chirally symmetric but confined phase that appears for quark chemical potentials larger than about a third of the nucleon mass, $\mu\gtrsim M_N/3$, can presumably be identified with the quarkyonic phase, that has been first proposed in~\cite{mcle07} and further studied in \cite{quarky} in particular. 

In the present work, we propose to re-build a PNJL model valid at large-$N_c$, but in which the Polyakov-loop potential is explicitly Z$_{N_c}$ symmetric. A possible way to answer to this requirement is to introduce a term in $\phi^{N_c}+\phi^{*N_c}$ instead of the standard Z$_3$-symmetric term $\phi^3+\phi^{*3}$. Such a potential is proposed in Sec.~\ref{pureglu}, and the corresponding PNJL model is written in Sec.~\ref{PNJLs}. Then, the issue of deconfinement and chiral symmetry restoration when varying $T$ and $\mu$ are discussed in Sec.~\ref{ccs} in 't Hooft's large-$N_c$ limit. The obtained phase diagram and concluding comments are given in Sec.~\ref{conclu}.

\section{Pure gauge sector}\label{pureglu}
\subsection{Explicit Z$_{N_c}$-symmetry}
The simplest effective potential energy density depending on $\phi$, defined in (\ref{phi}), and being explicitly Z$_{N_c}$-invariant has been proposed in~\cite{sanni05} and reads 
\begin{equation}\label{V1}
V_g(T,N_c,\phi,\phi^*)=A(T,N_c)\, |\phi |^2+B(T,N_c)\, |\phi |^4+C(T,N_c)\, (\phi^{N_c}+\phi^{*N_c}).
\end{equation}  
It is formally valid for any value of $\phi$, but one may restrict oneself to $|\phi |\in [0,1]$ in a mean-field approximation.  The above expression contains the basic blocks that could be expected to build a nontrivial theory: A mass term ($|\phi |^2$), an interaction term ($|\phi |^4$), and the term in $\phi^{N_c}+\phi^{*N_c}$ accounting for the explicit Z$_{N_c}$-symmetry \cite{comment}. Terms scaling like $|\phi |^6$, $|\phi |^8$, $\dots$, $|\phi |^{N_c-2}$, etc. could be added, but then the number of arbitrary functions would become too large to be efficiently constrained. Moreover, such higher-order terms would mostly be interaction terms that are already present in their simplest form in the $|\phi |^4$ term. The expression (\ref{V1}) is thus particularly convenient since it contains the minimal number of terms needed to perform the present study. The real coefficients $A$, $B$, and $C$ appearing in (\ref{V1}) are functions of $T$ and $N_c$ , and their explicit form will be specified in the following. Note that $\phi$, which depends on $T$, $N_c$, and $\vec x$ a priori, is here assumed to be independent of $\vec x$.  Beyond the polynomial form (\ref{V1}), logarithmic shapes can actually be shown to emerge from a Haar integration on the gauge group in a strong coupling expansion. One can find such a form in \cite{fuku03}, or for example in \cite{roessner}, where a potential schematically given by ${\cal U}/T^4=A(T) |\phi |^2+B(T) \ln\left[ 1-6|\phi |^2+4 (\phi^3+\phi^{*3})-3|\phi |^4\right] $ is used for $N_c=3$ computations. Instead of computing a similar potential at arbitrary $N_c$, we keep the ansatz (\ref{V1}) in the following; it is indeed particularly convenient for the calculations that are to be performed and still contains the  Z$_{N_c}$ symmetry we want to take into account. 

Various parametrizations of Z$_3$-symmetric potentials, fitted on pure gauge lattice data, have been proposed so far~\cite{polo82,fuku03,ratti06}.  Here, we are rather interested in obtaining an effective potential valid at large $N_c$, \textit{i.e.} $N_c>4$ at least. As shown below, all these values of $N_c$ will have in common that, in our approach, the``asymptotic" behaviour of $V_g$ (\textit{i.e.} values of $|\phi|$ larger that the physical one minimizing the potential) will be driven by the $Z_{N_c}$-symmetric term. The present formalism will thus not be valid for $N_c=3$ in particular, where this asymptotic behaviour is driven by the interaction term. The following expected qualitative behaviours have to be imposed in order to constrain the shape of the functions $A$, $B$, and $C$:
\begin{itemize}
\item The pressure $p_g=-{\rm min}_{\phi}(V_g)$ is proportional to $N_c^2\, T^4$ at large $N_c$ and $T$ in order to recover asymptotically the Stefan-Boltzmann limit of a free gluon gas. 
\item The norm, $|\phi_0|$, of the optimal value of the Polyakov loop, $\phi_0=|\phi_0|\, {\rm e}^{i\delta_0}$, is $N_c$-independent at the dominant order, see the definition (\ref{phi}). The first corrections, scaling as $1/N_c^2$, are neglected in the present approach -- more results on large-$N_c$ features of Wilson and Polyakov loops can be found for example in \cite{make}. $|\phi_0|=0$ in the confined phase, and $>0$ in the deconfined phase. Also, $|\phi_0|$ tends toward unity at very large $T$. 
\item There exists a critical temperature $T_d$ signalling a first-order phase transition, \textit{i.e.} the potential must have two different minima whose depth changes with the temperature in order to modify discontinuously the localisation of the absolute minimum. At the critical temperature, $|\phi_0|=0$ and $1/2$ are two degenerate minima of $V_g$. This last value is chosen so that it will ensure a good compatibility between our model and existing lattice data but it has only to be nonzero in order to lead to a deconfined phase. $T_d$ has to be seen as a typical value for the deconfinement temperature in SU($N_c$) Yang-Mills theory since the deconfinement temperature appears to be $N_c$-independent up to corrections in $1/N_c^2$~\cite{TcN,braun}.
\end{itemize} 
Obviously, the above constraint does not apply to $N_c=2$, where the transition is of second-order. This is not problematic since we eventually look for a model valid at large-$N_c$. Moreover,  the value  $|\phi_0|=1/2$ may not be the exact value of the Polyakov loop in $T_d$: Recent lattice results find it to be around 0.4 \cite{gupta07}, while a more recent renormalization-group-based approach leads to values closer to 0.6 for the Polyakov loop at the deconfinement temperature~\cite{braun}. The value 1/2 then appears to be relevant because it falls in the typical range of the existing results and because it simplifies the calculations performed in the following. 

The above constraints are actually satisfied by the following Lagrangian 
\begin{equation}\label{V2}
V_g=N_c^2 T^4\, a(T)\left[ |\phi|^2-4|\phi|^4+\frac{l(T)^{2-N_c}}{N_c}[8 l(T)^2-1](\phi^{N_c}+\phi^{*N_c})\right],
\end{equation}
where 
\begin{equation}\label{V2c}
a(T)>0,\quad l(T)>\frac{1}{\sqrt 8},\quad l(T_d)=\frac{1}{2},\quad  \partial_T l(T)> 0,\quad l(\infty)=1.
\end{equation}
Explicit forms of $a(T)$ and $l(T)$ will be given in the next section. All these conditions are required in order to have the existence of 2 degenerate minima and the correct behaviour of the Polyakov loop in the mean field approximation. The potential (\ref{V2}) has the following absolute minimum: $\phi_0(T<T_d)=0$ and $\phi_0(T\geq T_d)=|\phi_0(T)|\, {\rm e}^{2i\pi k/N_c}$, where $k=0,\dots,N_c-1$ and where $|\phi_0(T)|$ is a solution of
\begin{equation}\label{eq1}
1-8|\phi_0(T)|^2+l(T)^{2-N_c}\left[ 8l(T)^2-1\right] |\phi_0(T)|^{N_c-2}=0.
\end{equation}
It is straightforwardly checked that 
\begin{equation}
|\phi_0(T)|=l(T)
\end{equation}
actually solves (\ref{eq1}).

A more compact expression for the optimal value of the Polyakov loop is thus
\begin{equation}\label{V30}
\phi_0=l(T)\, {\rm e}^{2i\pi k/N_c}\, \Theta(T-T_d),
\end{equation}
where $\Theta$ is the Heaviside function. As seen from (\ref{eq1}), $|\phi_0|$ only depends on $T$ as required. 

\begin{figure}[t]
\includegraphics[width=8.5cm]{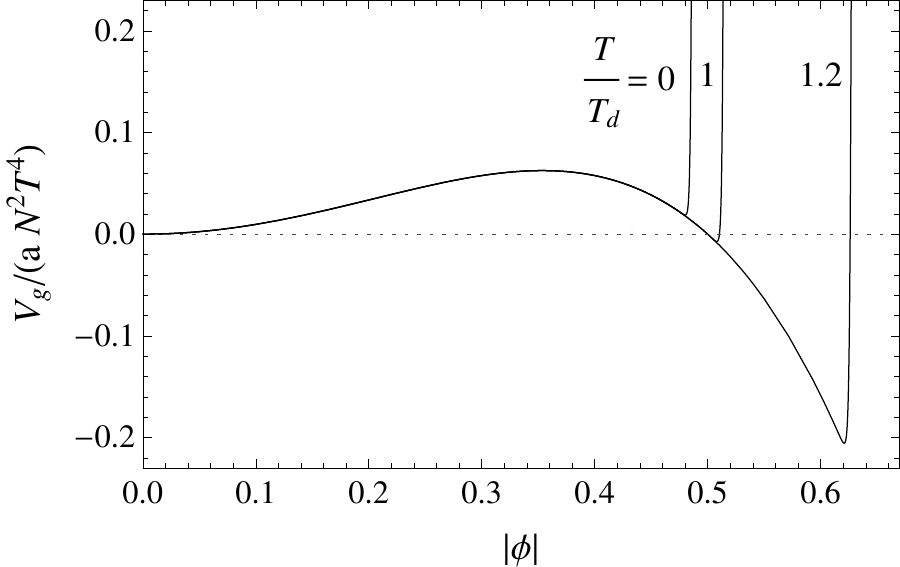}
\caption{Schematic evolution of the effective potential (\ref{V3}) versus the temperature (solid lines). }
\label{fig2}
\end{figure}

Restricting ourselves to the values $\phi=|\phi|\, {\rm e}^{2i\pi k/N_c}$, we get at the limit $N_c\rightarrow\infty$ a quite simple shape for the effective potential (\ref{V2}), namely
\begin{eqnarray}\label{V3}
\frac{V_g}{N_c^2 T^4}\equiv\frac{\omega_g}{T^4}&= \, a(T)\, |\phi |^2(1-4 |\phi |^2)  &\qquad |\phi |\leq l(T),\nonumber\\
&\rightarrow  +\infty &\qquad |\phi |>l(T).
\end{eqnarray}  
Hence, a U(1) invariance is recovered at infinite $N_c$ as a limiting case of the Z$_{N_c}$-symmetry. The schematic evolution of the large-$N_c$ limit of $V_g$ with the temperature is plotted in Fig.~\ref{fig2}; the behaviour (\ref{V3}) is readily observed, as well as the change of global minimum in $T=T_d$. 
Finally, the large-$N_c$ limit of the pressure reads
\begin{equation}\label{pg}
p_g(T,N_c)=N_c^ 2 T^ 4\, a(T)\, l(T)^2\left[ 4l(T)^2-1\right]. 
\end{equation}
Provided that $l(\infty)=1$ according to the large-$T$ behaviour of the Polyakov loop, $p_g$ would tend toward the Stefan-Boltzmann limit for a free gluon gas if $a(\infty)=\pi^2/135$.

\subsection{Numerical data}

\begin{figure}[t]
\includegraphics[width=8.5cm]{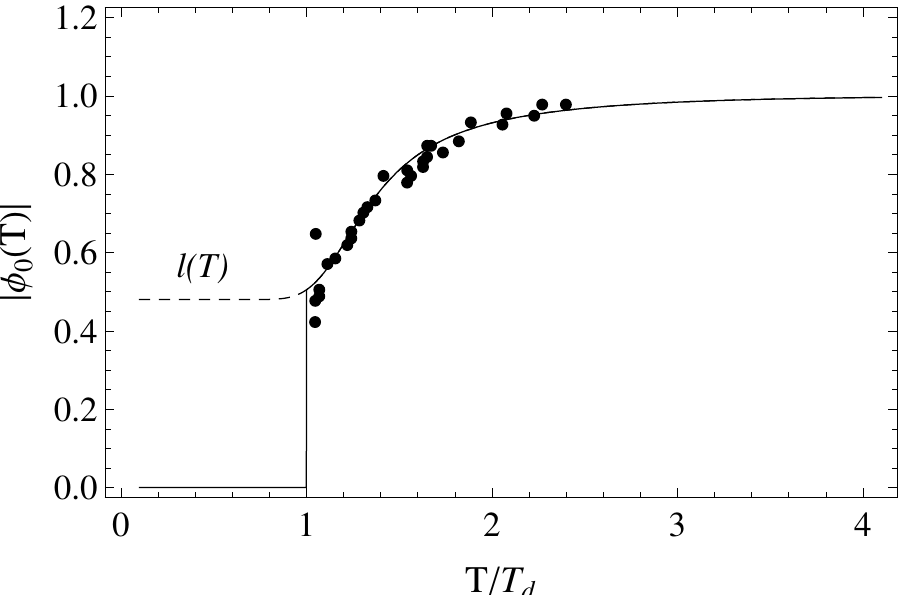}
\caption{Norm of the Polyakov loop minimizing the potential (\ref{V2}) versus the temperature in units of $T_d$ (solid line). The function $l(T)$ (dashed line) and the norm of the Polyakov loop computed in pure gauge SU(3) lattice QCD (points) have been added for comparison. Lattice data are given for temperatures lower than 2.4 $T_d$; data are taken from~\cite{gupta07}.}
\label{fig3}
\end{figure}

The function $l(T)$ is constrained by the relations (\ref{V2c}) in order for the structure of the potential and its evolution with the temperature to have the required behaviour. Moreover, $l(T)$ is equal to the norm of the Polyakov loop as soon as $T>T_d$. Those physical constraints are not sufficient to write down an explicit expression for $l(T)$. A possible way of proceeding, that we choose here, is to fit $l(T)$ on available lattice computations of the Polyakov loop in pure Yang-Mills theory. To our knowledge, large-$N_c$ values have not been obtained so far, but accurate SU(3) ones have been computed in~\cite{gupta07}. Since the Polyakov loop should not depend on $N_c$ at the dominant order, it is relevant to fit $l(T)$ on SU(3) data; the ad hoc form  
\begin{equation}\label{ldef}
l(T)=0.74-0.26\, \tanh\left[2.10\left(\frac{T_d}{T}\right)^3-0.60\frac{T}{T_d}\right]
\end{equation}
leads to a satisfactory parametrization of the results of \cite{gupta07} as it can be seen in Fig.~\ref{fig3}. It is also worth noting that Fig.~\ref{fig2} has been obtained using the form (\ref{ldef}) for $l(T)$. 

It is important to remark at this stage that the calculations we will perform are done in the mean-field approximation. In this scheme, the Polyakov loop is always lower than 1: values larger than 1 are due to quantum fluctuations and are \textit{de facto} beyond the mean-field treatment. That is why we have restricted our fit to lattice data lower than unity ($T<$ 2.4 $T_d$). We miss the overshoot due to quantum fluctuations, but we stay coherent with the mean-field approximation, and reach moreover $l(\infty)=1$. As a consequence, our results should be mostly trusted below $T<$ 2.4 $T_d$ but this is not a flaw since, in the following, we will be concerned with the phase structure of the theory and no phase transition will appear at energy scales above this upper limit. 

\begin{figure}[t]
\includegraphics[width=8.5cm]{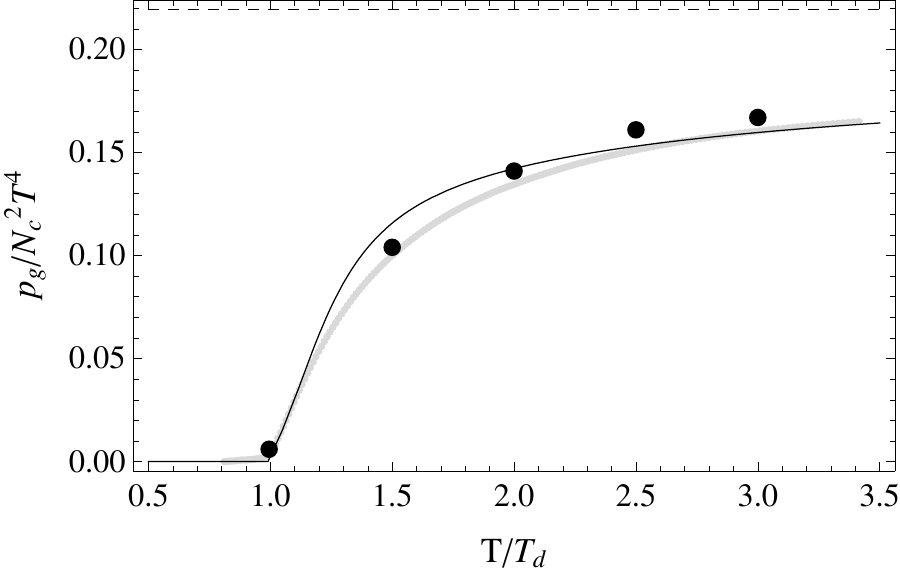}
\caption{Large-$N_c$ pure gauge pressure computed from Eq.~(\ref{pg}) and normalized to $N_c^2T^4$ (solid line). The corresponding lattice data, taken from \cite{pane09}, are plotted for comparison in the case $N_c=3$ (gray points) and $N_c\rightarrow\infty$ (black points).}
\label{fig4}
\end{figure}

The positive-definite function $a(T)$ is only present as an overall factor in $V_g$, so it does not come into play in the qualitative features of the effective potential. However, it is relevant in view of reproducing the absolute value of the pressure in pure gauge QCD, for which lattice data are known at $N_c=$ 3, 4, 5, 6, 8 and $\infty$ through an extrapolation of these data \cite{pane09}. The empirical choice
\begin{equation}\label{adef}
a(T)=\frac{1}{l(T)^4}\left(\frac{\pi^2}{135} -\frac{0.029}{\ln(T/T_d+1.5)}\right)
\end{equation}
leads to a good agreement between the lattice data of \cite{pane09} and formula (\ref{pg}), as shown in Fig.~\ref{fig4}. Notice that the value $a(\infty)$ is such that the Stefan-Boltzmann limit is reached at large temperatures. 

It is worth summarizing what has been done at this stage. Starting from Lagrangian (\ref{V1}), we have shown that the  three arbitrary functions of $T$ and $N_c$ it contains can be strongly constrained by demanding that the averaged Polyakov loop and the pure gauge pressure have a relevant behaviour in $T_d$, at large $T$, and in the large-$N_c$ limit. Explicit forms for the two remaining unconstrained functions of $T$ can then be found by asking the present model to be in agreement with current pure gauge lattice data. One is finally left with a fully determined Lagrangian with explicit Z$_{N_c}$ symmetry at finite $N_c$ and U(1) symmetry in the large-$N_c$ limit. This Lagrangian is obviously not predictive concerning the thermodynamics of the pure gauge sector, just as previously used Lagrangians like (\ref{Ud}). However, its knowledge is a necessary step in view of making predictions concerning the quark sector,  whose inclusion is discussed in the next section. 


\section{PNJL model}\label{PNJLs}

As shown in \cite{fuku03}, a minimal coupling of the NJL Lagrangian (\ref{njlL}) to a gauge field of the form  $A_\mu=A_0\, \delta_{\mu0}$ makes eventually appear the Polyakov loop in the quark grand potential. In the mean field approximation, one is led indeed to the quark potential~\cite{fuku03}
\begin{eqnarray}\label{Vq}
\frac{V_q(\mu,T,\sigma,L,L^\dagger)}{N_c\, N_f}&=&\frac{\sigma^2}{2g}-2\int \frac{d^3p}{(2\pi)^3} \times\\
&&\left\lbrace  E_p+\frac{T}{N_c}{\rm Tr}_c{\rm ln}\left[ 1+L\, {\rm e}^{-(E_p-\mu)/T}\right]+\frac{T}{N_c}{\rm Tr}_c{\rm ln}\left[ 1+L^\dagger\, {\rm e}^{-(E_p+\mu)/T}\right] \right\rbrace \nonumber,
\end{eqnarray}
where the Polyakov loop $L$ has been defined in (\ref{L}). In the above equality, 
\begin{equation}
E_p=\sqrt{p^2+(m_q-\sigma)^2}
\end{equation}
is the quark dispersion relation, with $m_q$ the quark bare mass and $\sigma$ related to the chiral condensate as follows
\begin{equation}\label{qq0}
\sigma= G \left\langle \bar q\, q\right\rangle.
\end{equation}
The coupling $G$ has to scale as $(N_c\, N_f)^{-1}$ in order for the potential (\ref{Vq}) to scale as $N_c\, N_f$, so it is convenient to define the coupling $g$ as
\begin{equation}
 g=G\, N_c\, N_f .
\end{equation}
Although $N_f$ is a priori arbitrary, our results are mostly valid at $N_f=2$. For higher values of $N_f$ indeed, the axial anomaly (not present in this formalism) should be taken into account in order to get a reliable model.  In what follows, $N_f=2$ will be implicitly understood, although we keep the notation $N_f$ so that the quark contributions appear more clearly. 

Since the pure gauge part of the potential only involves the traced Polyakov loop $\phi$, it is interesting to express $V_q$ in terms of $\phi$ rather than $L$. Terms of the form ${\rm Tr}_c{\rm ln}\left[ 1+z\, L\right]$ can be expressed as functions of ${\rm Tr}_c L\propto\phi$, ${\rm Tr}_c L^2$, ${\rm Tr}_c L^3$, \dots through a Taylor expansion. A possible way of proceeding is to expand the quark potential at the first order in $L$. This eventually leads to formulas in which only $\phi$ appears in $V_q$~\cite{mcle09}. This scheme has the advantage of being independent of the parametrization of $L$. Here we adopt an inequivalent procedure. As a first step, we notice that there exists in general a gauge in which the Polyakov loop $L$ is a diagonal element of $SU(N_c)$:
\begin{equation}
L={\rm diag}({\rm e}^{i\theta_1},{\rm e}^{i\theta_2},\dots ,{\rm e}^{i\theta_{N_c-1}},{\rm e}^{-i\sum^{N_c-1}_{j=1}\theta_j}).
\end{equation}
The $N_c-1$ parameters $\theta_j$ are real so that $L^\dagger L=\textbf{1}$ and ${\rm det}L=1$ as demanded for an SU($N_c$) element. In the special case of $N_c=3$, there is a one-to-one correspondence between the parameters $\theta_1$, $\theta_2$ and the Polyakov loop degrees of freedom $\phi$, $\phi^*$. This is not the case at large $N_c$ however, where the number of independent parameters in the Polyakov loop goes to infinity. As a consequence, an exact computation of the color traces appearing in (\ref{Vq}) is not possible unless simplifying assumptions are made. As a second step to reach this goal, we propose the following ansatz:
\begin{eqnarray}\label{Ldef}
L&={\rm diag}(\underbrace{{\rm e}^{i\theta},\dots ,{\rm e}^{i\theta}}_{(N_c-1)/2},1,\underbrace{{\rm e}^{-i\theta}, \dots,{\rm e}^{-i\theta}}_{(N_c-1)/2})&\qquad {\rm odd}-N_c\\
&={\rm diag}(\underbrace{{\rm e}^{i\theta},\dots ,{\rm e}^{i\theta}}_{N_c/2},\underbrace{{\rm e}^{-i\theta}, \dots,{\rm e}^{-i\theta}}_{N_c/2})& \qquad {\rm even}-N_c.\nonumber
\end{eqnarray}
It reduces to the mean-field parametrization of \cite{fuku03} at $N_c=3$, but the price to pay is that the number of degrees of freedom in $L$ is drastically reduced to a single real parameter $\theta$. It is then readily computed that
\begin{eqnarray}\label{phidef}
\phi&=\frac{1}{N_c}\left[1+(N_c-1)\cos\theta\right] &\qquad {\rm odd}-N_c \\
    &=\cos\theta &\qquad {\rm even}-N_c \nonumber
\end{eqnarray}
by using of the ansatz (\ref{Ldef}) in (\ref{phi}). We thus have an asatz that ``looks like" the SU(3) case and that reduces to $\phi=\cos\theta$ at large-$N_c$. 

Moreover, one can compute that 
\begin{eqnarray}
{\rm Tr}_c{\rm ln}\left[ 1+L\, {\rm e}^{-(E_p-\mu)/T}\right]&=&{\rm ln\, det}_c\left[ 1+L\, {\rm e}^{-(E_p-\mu)/T}\right]\nonumber\\
&=&\frac{N_c-1}{2}\, {\rm ln}\left[1+2\frac{N_c\phi-1}{N_c-1}{\rm e}^{-(E_p-\mu)/T}+{\rm e}^{-2(E_p-\mu)/T} \right] \nonumber\\
&&\ +{\rm ln}\left[1+{\rm e}^{-(E_p-\mu)/T}\right] \qquad {\rm odd}-N_c,  \\
&=&\frac{N_c}{2}\, {\rm ln}\left[1+2\phi{\rm e}^{-(E_p-\mu)/T}+{\rm e}^{-2(E_p-\mu)/T} \right] \nonumber\\
&& \qquad\qquad\qquad\qquad\qquad\quad\, {\rm even}-N_c,  \nonumber
\end{eqnarray}
and, taking into account a cutoff for the momentum integration of the vacuum term, one finally arrives at the quark potential, whose large-$N_c$ limit is given by
\begin{eqnarray}\label{Vq2}
\omega_q(\mu,T,\sigma,\phi)&=&\frac{V_q(\mu,T,\sigma,\phi)}{N_c\, N_f}\nonumber\\
&=&\frac{\sigma^2}{2g}-\frac{1}{\pi^2}\int^\Lambda_0 dp\, p^2  E_p-\frac{T}{2\pi^2}\int^\infty_0 dp\, p^2   \left\lbrace {\rm ln}\left[1+2\phi{\rm e}^{-(E_p-\mu)/T}+{\rm e}^{-2(E_p-\mu)/T} \right]\right. \nonumber\\ 
&&\left. +(\mu\rightarrow -\mu) \right\rbrace .
\end{eqnarray}
This last potential formally reduces to the genuine NJL potential once $\phi=1$, as observed in previous studies~\cite{fuku03,mcle09}. 
The total potential of the large-$N_c$ PNJL model under study is finally given by
\begin{equation}\label{potf}
{\cal V}(\mu,T,\sigma,\phi)=N_c^2\omega_g(T,\phi)+N_c\, N_f\, \omega_q(\mu,T,\sigma,\phi).
\end{equation}

In the confined phase, where $\phi=0$,  one observes a term in ${\rm ln}\left[1+{\rm e}^{-2(E_p-\mu)/T} \right]$ in the potential (\ref{Vq2}), so it could be tempting to associate such a term with diquark degrees of freedom in the confined phase. However, in the limit $T\rightarrow 0$, one exactly recovers the zero temperature NJL potential, expressed in terms of the quark degrees of freedom. So the confined degrees of freedom are still quarks in the present approach. Similarly, baryonic degrees of freedom (states with $N_c$ quarks) are not included in the present formalism; this would require further extensions of the present approach that are beyond the scope of the present study.

\section{Phase diagram at large $N_c$}\label{ccs}

In 't Hooft's large-$N_c$ limit, the number of quark flavours stays finite and ${\cal V}$ is dominated by the gluonic contribution. Consequently, when $N_c$ becomes infinite, the optimal value $\phi_0$ can be found by minimizing $\omega_g$ only.  According to (\ref{V30}), the large-$N_c$ solution (\textit{i.e} when $N_c$ is infinite) reads
\begin{equation}\label{phif}
\phi_0(T)=l(T)\, \Theta(T-T_d).
\end{equation}
The physical value of $\sigma$, denoted $\sigma_0$ and depending on $T$ and $\mu$, is then such that it minimizes $\omega_q(T,\mu,\sigma,\phi_0(T))$. $\omega_g$ does not depend on $\sigma$. Since $\sigma\propto\left\langle \bar q q\right\rangle$, chiral symmetry is present when $\sigma_0=0$ and broken when $\sigma_0\neq 0$. As a consequence of (\ref{phif}), the deconfined phase appears as soon as $T>T_d$, independently of the value of $\mu$: As pointed out in~\cite{mcle07}, quarks have no influence on the deconfinement phase transition at large-$N_c$ because of the suppression of internal quark loops in this limit. 

As a consequence of the large-$N_c$ limit, the confined/deconfined phases are straightforwardly identified in our model. The situation is less simple as far as chiral symmetry is concerned; numerical computations are needed. As a first step, the parameters of the model have to be fixed. The values
\begin{equation}\label{p1}
m_q=5.5~{\rm MeV},\  g=60.48~{\rm GeV}^{-2},\ \Lambda=651~{\rm MeV},\ T_d=270~{\rm MeV},
\end{equation}
used in the PNJL study \cite{ratti06}, will be taken in the following also. The first three parameters have been fitted so that the zero-temperature pion mass and decay constant are reproduced within the standard NJL model with $N_c=3$ and $N_f=2$ \cite{NJL,fit}. $T_d$ is a typical value for the deconfinement temperature in SU($N_c$) Yang-Mills theory. 

Using the parameters (\ref{p1}), the optimal value $\sigma_0$ can now be computed for any couple $(\mu,T)$, and can be linked to the quark condensate thanks to (\ref{qq0})
\begin{equation}
\left\langle \bar q q\right\rangle(\mu,T)=\frac{N_c\, N_f}{g}\sigma_0(\mu,T).
\end{equation}
In the limit where $T$ and $\mu$ both tend toward zero, we get
 \begin{equation}
\lim_{\mu,T\rightarrow 0}\left\langle \bar q q\right\rangle(\mu,T)=-N_c\, N_f\, 5.29\, 10^6\, {\rm MeV}^3,
\end{equation}
corresponding to a quite common value of $-$(317 MeV)$^3$ for $N_c=3$ and $N_f=2$. 

\begin{figure}[t]
\includegraphics[width=9.5cm]{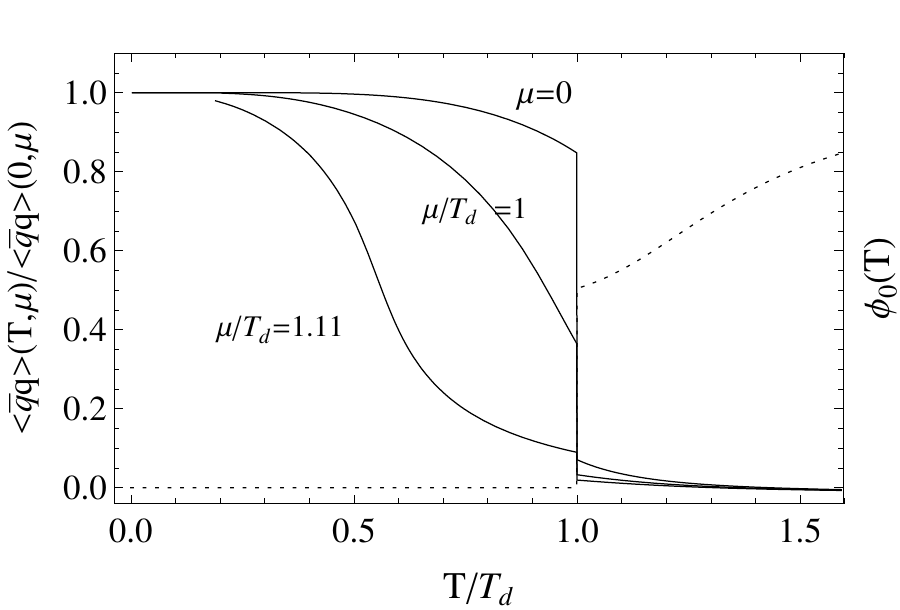}
\caption{Chiral condensate at large-$N_c$ ($N_c\rightarrow\infty$), normalized to its zero temperature value, versus $T$ in units of $T_d$, and plotted for $\mu/T_d=$ 0, 1, and 1.11 (solid lines). The optimal value of the Polyakov loop is also plotted (dotted line).}
\label{fig5}
\end{figure}

The large-$N_c$ chiral condensate versus the temperature is plotted in Fig.~\ref{fig5} for some values of the quark chemical potential. The most salient feature of this plot is the simultaneity of the first-order deconfinement phase transition and of the restoration of chiral symmetry through a first-order phase transition occurring at $T_\chi=T_d$. However, when $\mu/T_d\gtrsim 0.8$ ($\mu\gtrsim$ 200~MeV), the quick decrease of the chiral condensate suggests a progressive restoration of chiral symmetry through a crossover at temperatures smaller than $T_d$. As shown in~\cite{fuku03}, the crossover temperature can be computed thanks to the determination of the peak position in the dimensionless quark susceptibility reading, at large-$N_c$,
\begin{equation}\label{chiq}
\chi_{qq}(T,\mu)=\frac{\Lambda\, T}{\left.\partial^2_\sigma\omega_q\right|_{\sigma=\sigma_0}}.
\end{equation}
We have chosen to follow the definition of \cite{fuku03} for the quark susceptibility, although a more standard way of defining the susceptibility is rather  $\partial^2_{m_q}\omega_q(T,\mu)$, see \textit{e.g.} \cite{nonlo2}. In both cases, a peak in the quark susceptibility signals a phase transition . 

\begin{figure}[t]
\includegraphics[width=8.5cm]{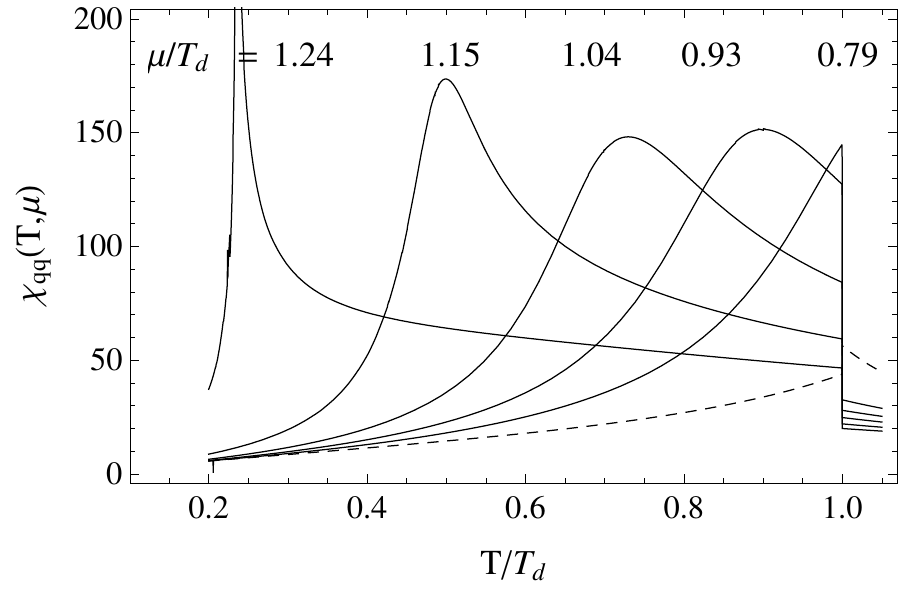}
\caption{Dimensionless quark susceptibility (\ref{chiq}) versus the temperature in units of $T_d$ (solid lines) with, from left to right, $\mu/T_d=$1.24, 1.15, 1.04, 0.93, 0.79. $\chi_{qq}(T,0)$ is also plotted for completeness (dashed line). Computations were done for $N_c\rightarrow\infty$. }
\label{fig4b}
\end{figure}

A plot of $\chi_{qq}(T,\mu)$ for some values of $\mu/T_d$ is given in Fig.~\ref{fig4b}. Several observations can be made by observing this figure together with Fig.~\ref{fig5}. First, the peak of the quark susceptibility is located in $T_d$ when $\mu/T_d\leq 0.79$; this corresponds to a first-order-type chiral symmetry restoration in the deconfined phase. The point $(0.79,1)\times\, T_d$ actually corresponds to a triple point in the $(\mu,T)$-plane: At large $\mu$ the peak of $\chi_{qq}$ is located below $T_d$ -- a larger $\mu$ corresponds to a lower peak position --, leading to the existence of a confined phase in which chiral symmetry is progressively restored through a crossover. A careful look at $\sigma_0$ actually shows that the chiral phase transition below $T_d$ becomes of first order when $\mu/T_d\geq 1.24$: There exists thus a critical-end-point that we find to be $(1.23,0.26)\times\, T_d$ in the $(\mu,T)$-plane. Apart from the susceptibility, the position of the chiral phase transition could have been alternatively determined by computing the zero of $\partial^2_T\sigma_0(T,\mu)$. We have checked that the chiral temperatures computed using that method agree with those computed with those computed from the peak in the susceptibility up to 5\%. For an exploratory study such as the present one, this agreement is satisfactory.  

\begin{figure}[t]
\includegraphics[width=8.5cm]{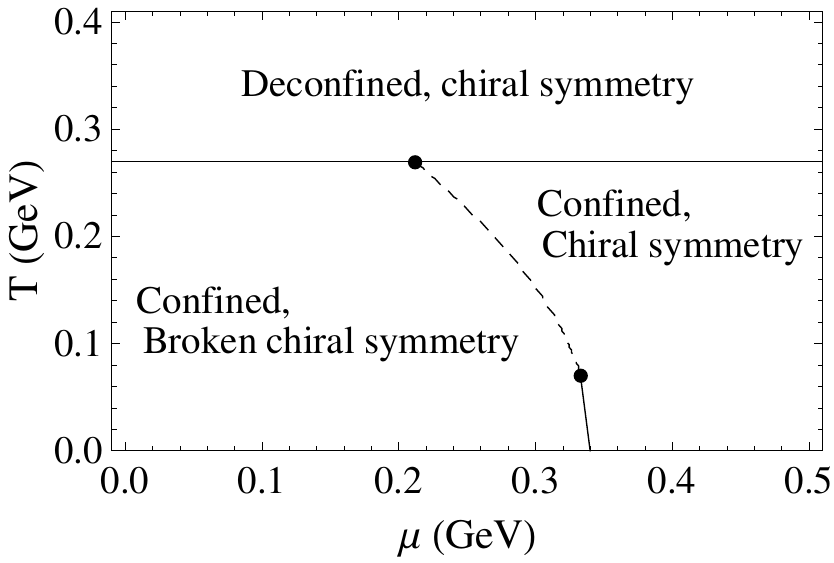}
\caption{Phase diagram of the large-$N_c$ PNJL model (\ref{potf}) with explicit Z$_{N_c}$ symmetry, obtained for $N_c\rightarrow\infty$. The solid lines denote first-order phase transitions while the dashed line denotes a crossover. The triple point (0.212, 0.270)~GeV and the critical end-point (0.335, 0.063)~GeV have been also plotted. The end of the lower curve is reached at (0.343, 0)~GeV.}
\label{fig6}
\end{figure}

Gathering all these observations, the phase diagram of our model in the $(\mu,T)$-plane can be established; it is shown in Fig.~\ref{fig6}. The three phases we find correspond to those found in~\cite{mcle09}, see Fig.~\ref{fig1}, but the structure of the chiral phase transition is a bit more involved under $T_d$: The chemical potential at which chiral symmetry is restored now depends on $T$, and there exists a critical-end-point at large enough $\mu$. Although the deconfining phase transition corresponds to what is expected in the large-$N_c$ limit of QCD from generic arguments~\cite{mcle07}, the critical line $T_\chi(\mu)$ we find under $T_d$ quite resembles to what can be observed within previously known $N_c=3$ PNJL studies~\cite{fuku03,ratti06}.  The similarity between our way of including the Polyakov loop in the NJL model and the way of \cite{fuku03}  --  our ansatz is a straightforward generalization of the one used in this last work -- might actually be at the origin of the similarities between the phase diagrams we find. The same reason, combined to the fact that we chose for our parameters values fitted on the SU(3) case, might explain why the values we find for \textit{e.g.} the chiral condensate are similar to those of \cite{fuku03}. 

We notice that, at large but finite values of $N_c$, the full potential (\ref{potf}) has to be minimized and quark contributions (presumably in $1/N_c$) will cause the Polyakov loop to be different from $l(T)$. Hence, the chiral condensate will also be modified, and the whole phase diagram will be affected. We nevertheless choose here to focus on the large-$N_c$ limit of the model, since it has been designed to be relevant in this limit mostly. 

\section{Conclusions}\label{conclu}

Effective ``Polyakov-loop-based" approaches have proven to be a relevant tool in view of modelling the thermodynamic properties of pure gauge QCD. The traced Polyakov loop is then the order parameter associated to confinement, itself seen as correlated with a global center symmetry, Z$_{N_c}$ when the gauge group is SU($N_c$). Following a suggestion made in~\cite{sanni05}, an explicitly Z$_{N_c}$-symmetric potential involving the traced Polyakov loop has been built and leads to: A first-order phase transition at large-$N_c$, a gluonic pressure scaling as $N^ 2_c$, and an $N_c$-independent optimal value for the Polyakov loop. The coupling of the pure gauge sector to light quarks has then been performed within a PNJL approach. Thanks to a particular ansatz for the Polyakov loop, the quark potential is such that it only involves the traced Polyakov loop that appears in the pure gauge potential. It has to be said that the resulting PNJL model is designed to be relevant in the large-$N_c$ limit only. It is moreover inequivalent to the model proposed in~\cite{mcle09} since the colour traces that have to be performed in the derivation of the PNJL potential are carried differently. 

The different assumptions that were made lead to a large-$N_c$ phase diagram that compares favourably with previous approaches. At any $\mu$, it shows a deconfined, chirally symmetric, phase above the deconfinement temperature ($T_d=$270~MeV). The deconfinement phase transition and the restoration of chiral symmetry are found to be simultaneous first-order phase transitions, in agreement with~\cite{mcle07,mcle09}. At temperatures lower than $T_d$, thus in the confined phase, there is a critical line $T_\chi(\mu)$ separating a phase with broken chiral symmetry at small $\mu$ and a chirally symmetric phase at large $\mu$. The phase transition is found to be a crossover from the triple point (0.212, 0.270)~GeV to the critical-end-point (0.335, 0.063)~GeV. It is then of first order until the boundary (0.343, 0)~GeV is reached, corresponding to the estimate $\mu\approx M_N/3$~\cite{mcle07}. It is worth saying that a confined, chirally symmetric, phase has also been found by solving Schwinger-Dyson equations at nonzero $\mu$ in~\cite{Gloz}, and that evidences for the existence of such a phase has been found in Coulomb gauge QCD calculations~\cite{swan}. The present model's prediction of three different phases, among which a chirally symmetric phase at baryonic chemical potentials around the nucleon mass thus seems a reliable one. The structure of the critical lines appears however to depend quite strongly on the way the Polyakov loop $L$ is handled: The comparison of Figs. \ref{fig1} and \ref{fig6} shows how a different treatment of the Polyakov loop within similar PNJL-based approaches can affect the phase diagram. It is also worth noting that an accurate description of the confined but chirally symmetric -- quarkyonic at large $N_c$ -- phase would require a more detailed approach that the simple PNJL model presented here. In particular, the explicit inclusion of baryonic degrees of freedom should be performed, but this is a task that we leave for subsequent studies. A first interesting attempt has for example been made in \cite{Cohen}, where a constituent approach is used to model the baryonic matter at large $N_c$. 

Recently, an effective model has been proposed \cite{sasaki}, in which the Lagrangian involves a linear sigma model for the quark part and a dilaton-like effective potential for the gluon part. Such a dilaton-like potential is designed to mimic the breaking of scale invariance through a nonzero value of the gluon condensate (the dilaton field). The phase diagram which is found is nearly identical to the one of Fig. \ref{fig6}, excepted that the deconfinement phase transition is replaced by a restoration of scale invariance. Both seem then to be linked, as suggested in \cite{sasaki}. 

Finally, we remark that the large-$N_c$ limit of the proposed pure gauge effective potential has a U(1) symmetry, which emerges as the limit of a Z$_{N_c}$ symmetry. This large-$N_c$ effective potential could be used in an approach where the traced Polyakov loop is allowed to depend on the position, typically via a Lagrangian of the type ${\cal L}\approx\partial_\mu\phi\, \partial^\mu \phi^*-V_g$. Of particular interest would then be to search for localised, solitonic, solutions of ${\cal L}$: One could then take advantage of the fact that finding solutions of a complex scalar field theory with a U(1) invariance is a topic that has attracted a considerable attention, mostly since Coleman's work on Q-balls~\cite{qb}, and for which many results are already available. We hope to present such a study in future works.  

\acknowledgments
G.L. thanks the F.R.S.-FNRS for financial support, and F. Sannino for comments at early stages of the manuscript.


\begin{thebibliography}{99}
\bibitem{comment} As pointed out in \cite{sanni05}, the term $\phi^{N_c}-\phi^{*N_c}$ possesses the same symmetry but violates charge conjugation, which is also a symmetry of the effective theory under consideration.
\end{thebibliography}
\end{document}